\documentclass[12pt, preprint]{aastex}
\usepackage{apjfonts, emulateapj5, natbib, color, CJK}
\input psfig.tex
\def\fnu{erg~s$^{-1}$~cm$^{-2}$~Hz$^{-1}$}
\newcommand{\flamb}{erg~s$^{-1}$~cm$^{-2}$ \AA$^{-1}$}
\newcommand{\flux}{erg~s$^{-1}$~cm$^{-2}$}
\newcommand{\lum}{erg~s\ensuremath{^{-1}}}
\newcommand{\lbol}{\ensuremath{L\mathrm{_{bol}}}}

\newcommand{\ledd}{\ensuremath{L\mathrm{_{Edd}}}}
\newcommand{\lratio}{\lbol/\ledd}

\newcommand{\msun}{\ensuremath{M_{\odot}}}
\newcommand{\lsun}{\ensuremath{L_{\odot}}}
\newcommand{\kms}{\ensuremath{\mathrm{km~s^{-1}}}}
\newcommand{\mbh}{\ensuremath{M_\mathrm{BH}}}

\newcommand{\chisq}{\ensuremath{\chi^2}}
\newcommand{\aox}{\ensuremath{\alpha_{\rm{ox}}}}

\newcommand{\ha}{H\ensuremath{\alpha}}
\newcommand{\hb}{H\ensuremath{\beta}}

\newcommand{\nii}{[N\,{\footnotesize II}]}
\newcommand{\sii}{[S\,{\footnotesize II}]}
\newcommand{\oiii}{[O\,{\footnotesize III}]}

\newcommand{\colnh}{\ensuremath{N_\mathrm{H}}}

\newcommand{\hst}{\emph{HST}}
\newcommand{\chandra}{\emph{Chandra}}
\newcommand{\xmm}{\emph{XMM-Newton}}
\newcommand{\galex}{\emph{GALEX}}

\newcommand{\sersic}{S\'{e}rsic}

\def\lax{{$\mathrel{\hbox{\rlap{\hbox{\lower4pt\hbox{$\sim$}}}\hbox{$<$}}}$}}
\def\gax{{$\mathrel{\hbox{\rlap{\hbox{\lower4pt\hbox{$\sim$}}}\hbox{$>$}}}$}}
\lefthead{Jiang et al.}
\righthead{UM~625: Sy1 with a low-mass BH}
\slugcomment{Accepted for publication in ApJ}

\begin{document}
\begin{CJK*}{UTF8}{gbsn}

\title{UM~625 Revisited: Multiwavelength Study of A Seyfert 1 Galaxy 
with a Low-mass Black Hole}
\author{{Ning~Jiang (蒋凝)\altaffilmark{1,2}}, Luis~C.~Ho\altaffilmark{2},
Xiao-Bo~Dong (董小波)\altaffilmark{1,3}, Huan~Yang (杨欢)\altaffilmark{1} 
and Junxian~Wang (王俊贤)\altaffilmark{1}}
\altaffiltext{1}{Key laboratory for Research in Galaxies and Cosmology, 
Department of Astronomy, The University of Science and Technology of China,
Chinese Academy of Sciences, Hefei, Anhui 230026, China; ~jnac@mail.ustc.edu.cn,
~xbdong@ustc.edu.cn}
\altaffiltext{2}{The Observatories of the Carnegie Institution for Science, 
813 Santa Barbara Street, Pasadena, CA 91101, USA; ~lho@obs.carnegiescience.edu}
\altaffiltext{3}{Yunnan Astronomical Observatory, Chinese Academy of Sciences, 
Kunming, Yunnan 650011, China; Key Laboratory for the Structure and Evolution of 
Celestial Objects, Chinese Academy of Sciences, Kunming, Yunnan 650011, China}

\begin{abstract}
UM~625, previously identified as a narrow-line active galactic nucleus (AGN), 
actually exhibits broad \ha\ and \hb\ lines whose width and luminosity 
indicate a low black hole mass of $1.6 \times 10^6$ \msun.  We present a 
detailed multiwavelength study of the nuclear and host galaxy properties of 
UM 625.  Analysis of \chandra\ and \xmm\ observations suggests that this 
system contains a heavily absorbed and intrinsically X-ray weak ($\aox=-1.72$) 
nucleus.  Although not strong enough to qualify as radio-loud, UM 625 does 
belong to a minority of low-mass AGNs detected in the radio.  The broad-band 
spectral energy distribution constrains the bolometric luminosity to 
$\lbol\approx(0.5-3)\times10^{43}$~\lum\ and $\lratio\approx0.02-0.15$.
A comprehensive analysis of Sloan Digital Sky Survey and {\it Hubble Space 
Telescope}\ images shows that UM~625 is a nearly face-on S0 galaxy with a
prominent, relatively blue pseudobulge (\sersic\ index $n = 1.60$) 
that accounts for $\sim$60\% of the total light in the $R$ band.  The 
extended disk is featureless, but the central $\sim150-400$~pc contains a 
conspicuous semi-ring of bright, blue star-forming knots, whose integrated 
ultraviolet luminosity suggests a star formation rate of $\sim$0.3 
\msun~yr$^{-1}$.  The mass of the central black hole roughly agrees with the
value predicted from its bulge velocity dispersion but is significantly 
lower than that expected from its bulge luminosity.
\end{abstract}


\section{Introduction}
Supermassive black holes (BHs), with masses in the range of
$10^{6}-10^{10}$~\msun\ as measured via stellar and gas kinematics, have been
convincingly inferred to be present in the centers of nearby inactive
massive galaxies and are generally believed to reside in all
galaxies with a spheroidal stellar component (see Kormendy \& Ho 2013 for
a review). Moreover, there are tight relations between the BH mass
(\mbh) and the properties of the spheroidal component (namely, ellipticals and 
the bulges of disk galaxies), including stellar velocity dispersion 
($\sigma_\star$; Ferrarese \& Merritt 2000, Gebhardt et al. 2000a), luminosity 
($L_{\rm bulge}$; Kormendy \& Richstone 1995; Magorrian et al. 1998; Marconi 
\& Hunt 2003), and mass ($M_{\rm bulge}$; H\"aring \& Rix 2004). 

However, the situation is far from clear in the low-mass regime 
($\mbh \lesssim 10^6$ \msun) because such BHs are largely beyond the reach of 
current capabilities for direct dynamical measurement. A practical approach 
is to search for them in low-luminosity type 1 active galactic nuclei (AGNs)
with mass estimated from their broad-line width and luminosity using empirical 
virial relationships (e.g., Gebhardt et al. 2000b; Kaspi et al. 2000).
This technique has yielded a sample of $\sim 200-300$ candidate 
BHs with masses between $10^5$ to $10^6$ \msun (Greene \& Ho 2004, 2007b; 
Dong et al. 2012b), including a couple below $10^5$ \msun, in the
regime of so-called intermediate-mass BHs (IMBH, $10^{3-6}$ \msun; 
Filippenko \& Ho 2003; Barth et al. 2004; Dong et al. 2007; see reviews in 
Ho 2008; Kormendy \& Ho 2013).  The host galaxies of low-mass BHs thus found 
appear very different from their supermassive counterparts.
The best nearby example, NGC~4395, is a dwarf Sdm galaxy without a bulge
at all (Filippenko \& Ho 2003), while the second prototype, POX~52, is instead
a spheroidal or dwarf elliptical galaxy (Barth et al. 2004; Thornton et al. 
2008). According to the studies to date, the \mbh--$\sigma_{\star}$ relation 
of local inactive massive galaxies appears to roughly extend to the low-mass 
end (Barth et al. 2005; Greene \& Ho 2006; Xiao et al. 2011), but the 
\mbh--$L_{\rm bulge}$ relation does not.  Photometric decomposition by 
Greene et al. (2008) and Jiang et al. (2011a) indicates that the majority of 
the host galaxies with disks are likely to contain pseudobulges; the rest
resemble spheroidals according to their position on the fundamental plane.
Very few live in classical bulges.  Moreover, the \mbh--$L_{\rm bulge}$ 
relation flattens out at the low-mass end, and on average the bulge luminosity 
is larger by 1--2 index at fixed \mbh\ (Greene et al. 2008; Jiang et al. 2011b).

The low BH masses and high accretion rates that generally typify the 
low-mass AGN sample also provide a unique opportunity to probe accretion 
processes in an under-explored regime of parameter space.  Especially 
interesting is the broad-band spectral energy distribution (SEDs) of these 
systems.  Unlike quasars, which have luminosities that far exceed that of 
their host galaxies, the SEDs of Seyferts, particularly those with low-mass 
BHs, can be strongly contaminated by emission from their hosts.  Observations 
with high angular resolution are absolutely essential to decouple the AGN from 
the host to construct proper nuclear SEDs.  In spite of these challenges,
studies so far already suggest that low-mass AGNs may possess unusual
multiband properties, including for the tendency to be very radio-quiet 
(Greene et al. 2006; Greene \& Ho 2007b) and relatively X-ray bright (Greene 
\& Ho 2007a; Miniutti et al. 2009; Desroches et al. 2009; Dong et al. 2012a).

The most extensive multiwavelength studies on low-mass BHs thus far have
concentrated, unsurprisingly, on the two prototypes, NGC~4395 and POX~52.
Interestingly, the SED of NGC 4395 differs markedly 
from those of both quasars and typical low-luminosity AGNs (Moran et al. 1999).
Specifically, the big blue bump, which dominates the SED of luminous 
\begin{figure*}[t]
\centerline{\psfig{file=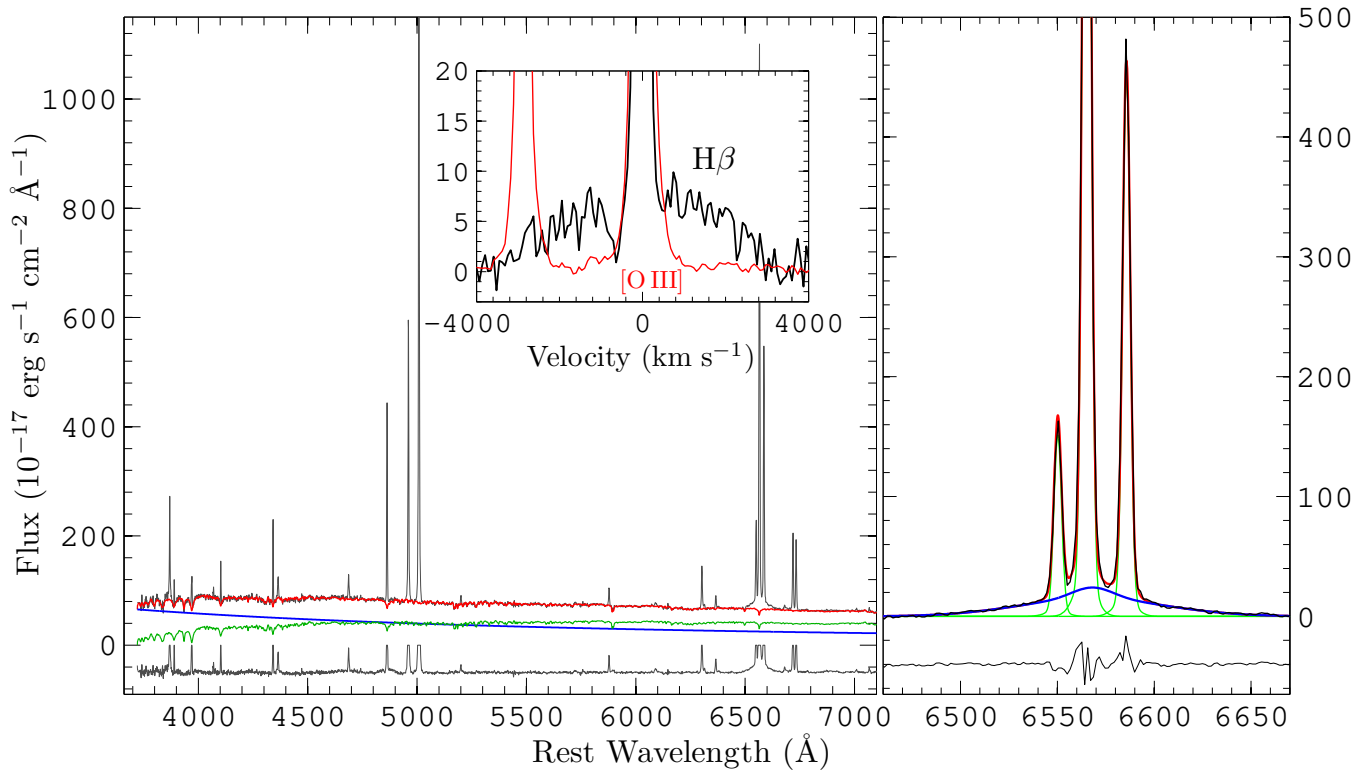,width=18cm}}
\figcaption[spec.eps]{
The SDSS spectrum of UM~625 (black),
together with its continuum decomposition and line fitting. ({\it Left})
The model for the starlight, featureless continuum, and the sum of the two
are plotted as green, blue, and red lines, respectively.  The residual
spectrum (observed $-$ model) is shown on the bottom, offset downward for clarity.
The inset highlights the \hb\ and [O~III] region
and the clear detection of broad \hb\ in the residual spectrum.
({\it Right}) Decomposition of
the H$\alpha +$[N II] region. The upper panel shows the original data
(thick black line), the fitted narrow lines (green), the fitted
broad H$\alpha$ component (blue), and the sum of all the fitted components
(red). The lower panel shows the residuals of the fit.}
\label{specfit}
\end{figure*}
\noindent
Seyferts and quasars at optical and UV bands, is absent, leading to
an optical-to-X-ray spectral index of $\aox=-0.97$ (Dewangan et al. 2008).
NGC~4395 is a rare example of a low-mass AGN with a relatively low Eddington 
ratio ($\lratio$) of $1.2\times10^{-3}$.  Most likely because of strong 
selection effects, 
optically selected samples of low-mass BHs tend to have higher $\lratio$, 
ranging from $\lesssim0.01$ to $\sim1$ with a median of 0.2 (Dong et al. 
2012b).  POX~52, with $L_{\rm bol}=1.3\times10^{43}$ \lum\ and $\lratio=0.2
-0.5$, exhibits a much more normal SED that is broadly similar to that of a 
scaled-down version of radio-quiet quasars (Thornton et al. 2008).
The broad-band SED of NGC~4051, a narrow-line Seyfert 1 galaxy with 
$\mbh=(1.7\pm0.5)\times10^6\msun$ (Denney et al. 2009) accreting at $1\%-5\%$ 
of \ledd, is best fit by a relativistically outflowing jet model 
(Maitra et al. 2011).

This paper reports a detailed multiwavelength data analysis of a low-mass BH 
hosted in the pseudobulge of UM 625.  First noted as a ``neutral compact 
spherical disc galaxy'' by Zwicky et al. (1975), the blue color of UM 625 
($B-V=0.67$ mag; Salzer et al. 1989) placed it in early catalogs of blue 
compact galaxies (e.g., Campos-Aguilar et al. 1993).  A high spatial 
resolution optical image was obtained with \hst\ as part of the nearby AGN 
survey of Malkan et al. (1998), in which UM~625 was noted as an S0 galaxy with 
a partial nuclear ring structure. Later analysis of a near-infrared (NIR) 
\hst/NICMOS image revealed a point-like nucleus embedded in an exponential 
surface brightness profile (Quillen et al. 2001; see also Hunt \& Malkan 2004).
An \hst\ ACS/HRC image taken in the near-ultraviolet (NUV) shows a very bright, 
compact, yet partially resolved nucleus (Mu{\~n}oz Mar{\'{\i}}n et al. 2007).

An optical spectrum of UM~625 was first acquired as part of the University of
Michigan objective-prism survey for emission-line galaxies (MacAlpine \& 
Williams 1981), from which its common name originates.  Since then 
UM~625 has been classified as a Seyfert 2 galaxy (or occasionally as an H II 
galaxy) in light of its extremely strong \oiii~$\lambda\lambda 4959,5007$ lines
and the apparent absence of broad emission lines (e.g., Salzer et al. 1989; 
Terlevich et al. 1991).  The spectral classification was re-examined 
critically by Dessauges-Zavadsky et al. (2000); using emission-line 
measurements from the literature, they classified it as a Seyfert 2 according 
to the three standard optical line-ratio diagnostic diagrams of Veilleux \& 
Osterbrock (1987).  Using a high-quality Sloan Digital Sky Survey (SDSS; York 
et al. 2000) spectrum with a resolution
$R \approx 2000$, we clearly detect both broad \ha\ and \hb\ and revise 
the spectral classification of UM 625 to a Seyfert 1.  Combining the broad 
line width and luminosity, we estimate a BH mass of $1.6 \times 10^{6}$ \msun,
placing UM~625 in our sample of AGNs with low-mass BHs (Dong et al. 2012b).

We assume a cosmology with $H_{0} =70$ km~s$^{-1}$~Mpc$^{-1}$, $\Omega_{m} = 
0.3$, and $\Omega_{\Lambda} = 0.7$.  At a redshift of $z=0.0250$, UM 625 
has a luminosity distance of 109.1 Mpc. 

\section{Analysis of the Optical Spectrum}

UM~625 was spectroscopically observed by SDSS on 18 June 2002 UT with an 
exposure time of 4803~s exposure.  It was classified as a galaxy by the 
spectroscopic pipeline of the SDSS Fourth Data 
Release (Adelman-McCarthy et al. 2006).  In the course of our systematic 
spectral fitting of all extragalactic objects in the SDSS, we first noted 
UM~625 because it showed evident broad \ha\ and \hb\ emission; its inferred 
virial BH mass placed it in the sample of type 1 AGNs with BH masses below 
$2 \times 10^6$~\msun\ described by Dong et al. (2012b).  The details of 
the spectral analysis are given in Dong et al. (2012b). Here we present
a brief description of the continuum modeling and emission-line profile 
fitting, which are based on the MPFIT package (Markwardt 2009) that performs 
\chisq-minimization by the Levenberg--Marquardt technique.

UM~625 has a redshift of $z=0.0250$, and its SDSS spectrum is dominated by
host galaxy starlight. The median signal-to-noise (S/N) ratio in the 
\hb--\oiii\ and \ha--\nii--\sii\ regions are 45 and 60 pixel$^{-1}$, 
respectively, high enough to fit accurately the continuum and emission lines.
We begin by correcting the spectrum for Galactic extinction using the extinction
map of Schlegel et al.\ (1998) and the reddening curve of Fitzpatrick (1999).
We model the starlight component with the stellar templates of Lu et al. (2006),
which were built from the simple stellar population spectra (Bruzual \& 
Charlot 2003).  The AGN continuum is modeled as a power law.  The stellar 
absorption lines must be subtracted well to ensure reliable measurement of 
weak emission lines (e.g., Ho et al. 1993, 1997). This is achieved by 
broadening and shifting the starlight templates to match the stellar velocity 
dispersion of the galaxy.  As shown in Figure~1 (left panel), 
the fit is good; the absorption features are well matched, and the residuals 
in the emission line-free regions are consistent with the noise level.  

Next, we fit the emission lines with Gaussians, using the code described in 
detail in Dong et al. (2005).  The spectrum of UM~625 is dominated by 
strong, narrow emission lines such as \oiii\,$\lambda\lambda4959,5007$, \hb,
\ha, and \nii\,$\lambda\lambda6548,6583$.  But even a cursory inspection of 
the spectrum reveals that \ha\ has a clear broad component, even before 
continuum subtraction; after continuum subtraction, a broad component to 
\hb\ also emerges.  This is reminiscent of the situation of the two prototypal 
broad-line AGNs with intermediate-mass BHs, namely NGC\,4395 (Filippenko \& 
Sargent 1989) and POX\,52 (Barth et al. 2004).  Because of their narrowness as 
well as the high S/N ratio of the SDSS spectrum, the doublet lines of both 
\sii\ and \nii\ are well isolated.  \sii, \nii, and the narrow components of 
the Balmer lines have very similar full widths at half-maximum (FWHMs).
We fit \sii\ first, assuming that the doublet lines have the same width;
a good fit is achieved with reduced $\chisq =1.1$ (29 degrees of freedom)
by using two Gaussians for each line.  Then we fit the \ha--\nii\ region, 
taking the best-fit model of \sii\ as a template to model \nii\ and narrow \ha.
The line ratio of \nii\ $\lambda$6583/$\lambda$6548 is set to the theoretical 
value 2.96, and their separation is fixed to the laboratory value.  We use 
additional Gaussians to model the broad component of \ha, starting with one 
Gaussian and adding in more if the fit can be improved significantly according 
to the $F-$test.  A good fit is achieved for broad \ha\ with just two 
Gaussians (Figure~1, right panel).  As the broad \hb\ 
component is weak, we fit the total \hb\ profile by assuming that its narrow 
component has the same profile as \sii\ and that its broad component has the 
same profile as broad \ha.  With the exception of \oiii\,$\lambda4959,5007$,
\psfig{file=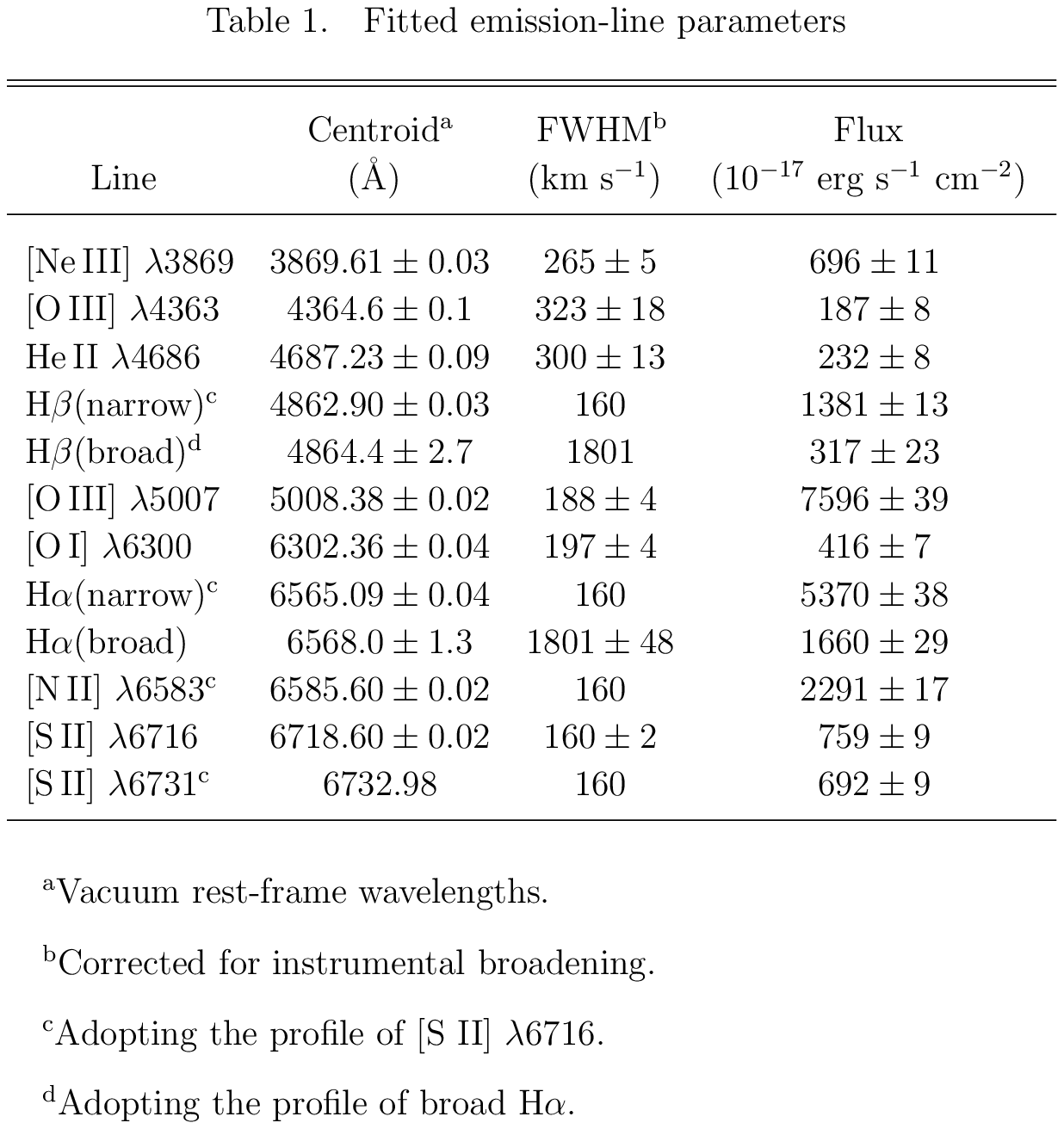,width=5.5cm,angle=0}
\vskip 0.1cm
\noindent
we fit the other narrow emission lines simply with a single Gaussian.  For \oiii,
we modeled each of its doublet lines with two Gaussians, one accounting for 
the bulk component (line core) and the other for a weak, yet apparent, blue 
wing.  All the line parameters are listed in Table 1.  We make available 
online the data and the detailed fitting parameters.%
\footnote{Available at
http://staff.ustc.edu.cn/\~{ }xbdong/Data\_Release/IMBH\_DR4/, together with 
auxiliary code to explain and demonstrate the fitting procedure.}

The ratios of the prominent narrow lines, \oiii\,$\lambda5007$/\hb\ $= 5.5$ 
and \nii\,$\lambda6583$/\ha\ $= 0.4$, place UM~625 within the regime of AGNs 
(Baldwin et al. 1981). With a total \hb\ to \oiii\,$\lambda5007$ ratio of 0.2, 
UM~625, like NGC 4395 and POX 52, would be considered a Seyfert 1.8 in the 
classification scheme of Osterbrock (1981).  From the best-fit model of broad 
\ha\ yields FWHM = $1801 \pm 68$~\kms\ and a line dispersion 
($\sigma_{\rm line}$, the second moment of the line profile) of 
$1583 \pm 59$~\kms\ (both corrected for the SDSS instrumental resolution of 
139~\kms\ FWHM).  The luminosity of broad \ha\ is $2.3 \times 10^{40}$ \lum.
The observed (not corrected for internal extinction) luminosity of the narrow 
\ha\ component is 3 times higher ($7.6 \times 10^{40}$ \lum), and the 
luminosity of \oiii\,$\lambda5007$ is $1.1 \times 10^{41}$ \lum.

To test the reliability of the broad components of Balmer lines, we refit the 
\ha--\nii\ complex and \hb\ using as a template for each line the 
\oiii\,$\lambda 5007$ profile derived from its double-Gaussian model.  The 
results are totally unacceptable; the reduced $\chisq = 12$ and 37, 
respectively, for the \hb\ and the \ha--\nii\ complex, with obvious large 
residuals.  This is evident by directly comparing the profiles of \hb\ and 
\oiii\,$\lambda5007$, as shown in the inset of Figure~1
(with \oiii\,$\lambda 5007$ scaled to have the same peak flux density as \hb);
\hb\ has an additional much broader, albeit low-contrast ($f_{\lambda} 
\lesssim 8 \times 10^{-17}$ \flamb), component that is not present in \oiii\,$\lambda 5007$.
Also evident in the inset is that \oiii\ is slightly broader than \hb\ 
[by $\sim120$ \kms, in height range $f_{\lambda} \approx (10-20) \times 
10^{-17}$ \flamb].  We also tried another fitting scheme in which we 
model the narrow Balmer lines with the total profile of \oiii\,$\lambda 5007$ 
and the broad component of the Balmer lines with two Gaussians;
the fitting results are also unacceptable, with reduced $\chisq = 7$ and 23, 
respectively, for the \hb\ and \ha--\nii\ regions.

The extinction of the broad-line region (BLR) can be derived
\begin{figure*}[t]
\centerline{\psfig{file=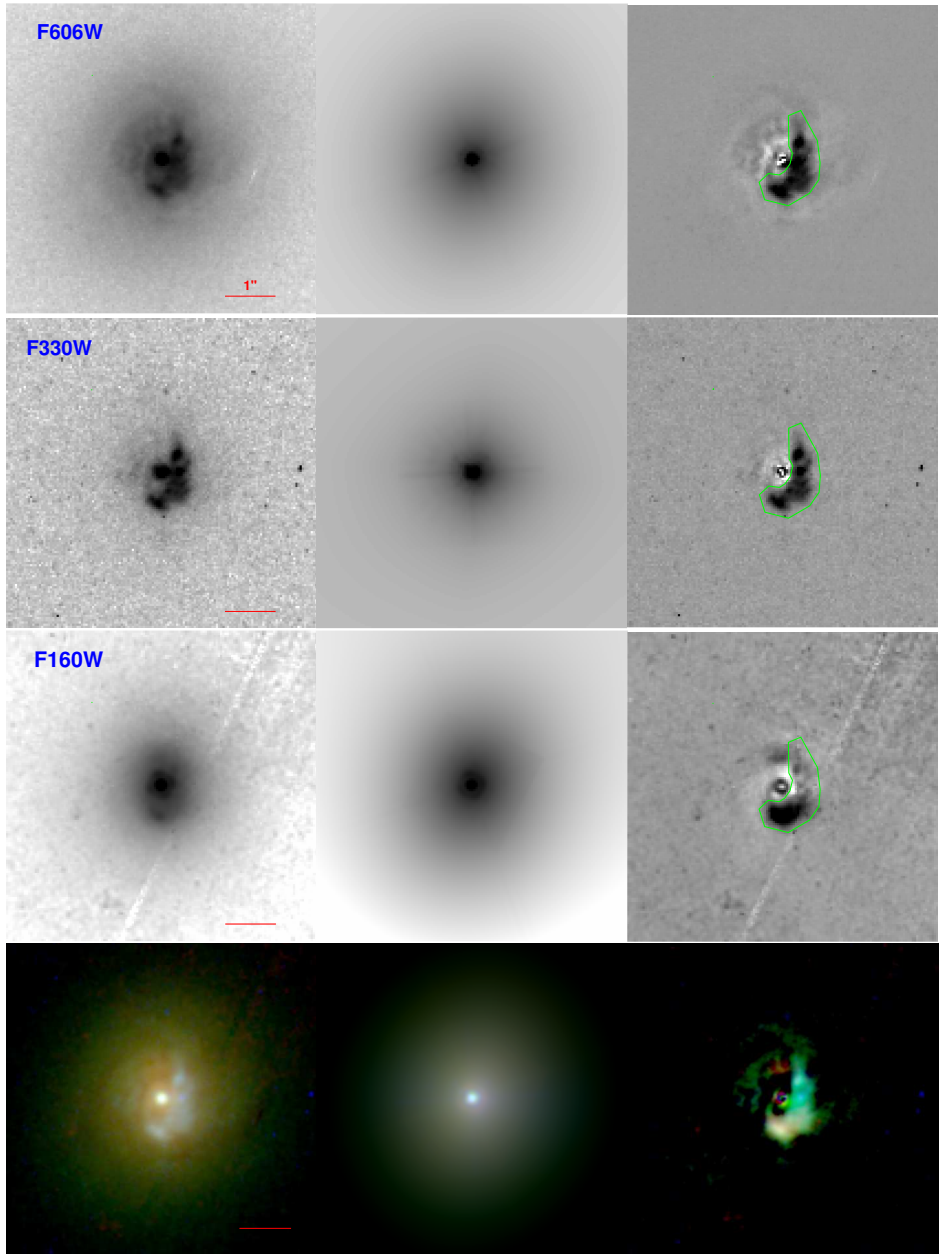,width=18cm}}
\figcaption[galfit.ps]{
The archival \hst\ images of UM~625 and 2-D imaging decomposition
by GALFIT. From top to bottom are F606W (roughly $R$ band;
WFPC2/PC1), F330W ($U$ band; ACS/HRC), F160W ($H$ band; NICMOS/NIC1)
and composite images of the three bands, respectively.  The left column shows
the original image, the middle column the GALFIT model (PSF + \sersic\ + disk),
and the right column the residual image.  All images are oriented with north
up and east to the left; the red line marks a scale of 1\arcsec\ ($\sim0.5$ kpc).
The green polygon denotes the star-forming ring region, which has been masked
out in the GLAFIT fitting.
}
\label{galfit}
\end{figure*}
from the observed 
Balmer decrement \ha/\hb\ for normal AGNs (Dong et al. 2008).  Assuming the 
extinction curve of the Small Magellanic Cloud (Hopkins et al. 2004; Wang et 
al. 2005) and an intrinsic broad-line \ha/\hb\ = 3.1 (Dong et al. 2008), we 
get $E(B-V)=0.57$ mag for the BLR.  Likewise, the observed
narrow \ha\ and \hb\ gives $E(B-V)=0.25$ mag for the NLR.  

\section{Analysis of the Images}

There are archival \hst\ images in three bands for UM~625: F330W (roughly the 
$U$ band of the Johnson system), F606W ($R$), and F160W ($H$), observed 
through ACS, WFPC2, and NICMOS, respectively.  From these images, we can see 
clearly that there is a bright point-like source present in the center of 
UM~625, and that the galaxy is almost round in shape and has no spiral arms, 
indicating an elliptical/spheroidal or a face-on S0 galaxy.  Closer inspection 
reveals that there is, in addition, a nuclear semi-ring on scales of $\sim 
150-400$ pc; this feature, most conspicuous in the F330W image, is resolved 
into several knots, presumably sites of intense star formation.  The ring 
region is indicated in Figure~2 with the green polygon.
These high-resolution images give us a distinct opportunity to study not only 
the multiwavelength properties of the AGN, but also its host galaxy.  The \hst\ 
images, however, suffer from one major limitation: their field of view is 
too small to properly sample the disk component or to reliably measure the 
sky background.  
\begin{figure*}[t]
\centerline{\psfig{file=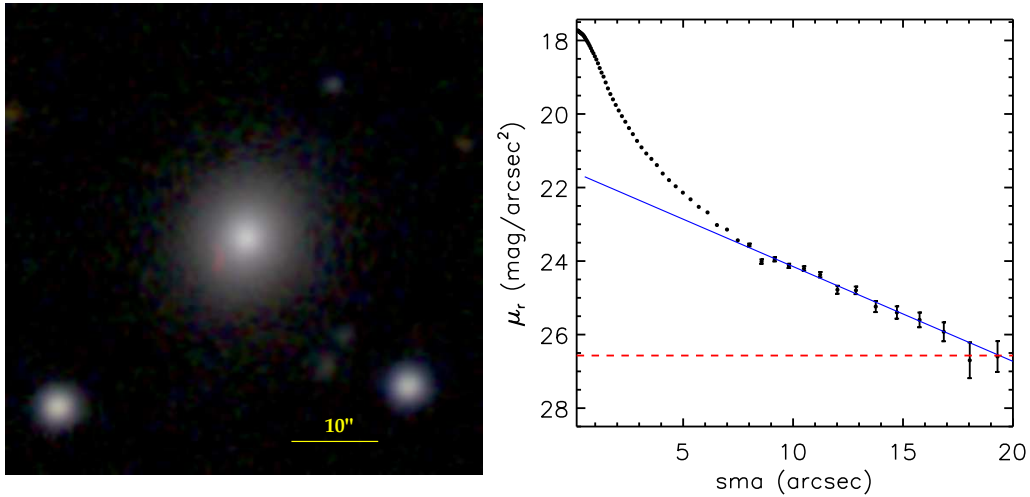,width=18cm}}
\figcaption[sdss.ps]{
({\it Left}) The SDSS three-color ($gri$) composite image of UM~625;
the yellow line marks a scale of $10\arcsec$.
({\it Right}) The $r$-band surface brightness profile ($\pm1\sigma$ error
bars) as a function of semi-major axis radius, with sky subtracted. The
red dotted line denotes the limiting surface brightness of
26.57~$\rm mag~arcsec^{-2}$.  We fit the profile at $r > 10$\arcsec\ with
an exponential disk model (blue line), which gives a scale length of
4\farcs21 (2.12~kpc).
}
\label{sdssimg}
\end{figure*}
\noindent

Fortunately, UM 625 was imaged with SDSS.  Thus, before we 
discuss the \hst\ images in detail, we first turn to the SDSS images. Our 
strategy is to use the SDSS $r$-band image, which closely approximates 
the \hst\ F606W band, to constrain the photometric parameters of the 
disk component, which is otherwise difficult to determine from the \hst\ 
data alone.  With the disk thus constrained, we will use the \hst\ F606W 
image to measure the parameters of the AGN point source and the bulge, which 
are central to our scientific analysis.  The F330W and F160W images provide 
further photometric points for the nuclear SED, as well as color information 
to diagnose the stellar population of the bulge.

\subsection{SDSS Imaging}

UM~625 was observed by SDSS in $ugriz$ on 24 May 2001 UT (Figure~3).
Although the standard SDSS images have a relatively short exposure time of 
only 54~s per filter, the drift-scan mode in which the imaging survey was 
conducted (Gunn et al. 1998) ensures very accurate flat-fielding.  Moreover, 
for most galaxies the field of view is quite large.  Both factors are crucial 
for obtaining an accurate measurement of the sky background and its associated 
error, which determines the limiting surface brightness sensitivity.
For most galaxies the azimuthally averaged surface brightness profile can be 
measured reliably down to $\mu_r \approx 27$~mag\,arcsec$^{-2}$,
which is deep enough to study the outer structure of galaxies
(Pohlen \& Trujillo 2006; Erwin et~al. 2008; Li et al. 2011).

We chose to work with the $r$ band, which is closest to the \hst\ F606W filter.
We first mask all photometric objects in the field identified either by 
Sextractor 
(Bertin \& Arnouts 1996) or by the SDSS photometric pipeline and then run 
the IRAF\footnote{IRAF (Image Reduction and Analysis Facility) is distributed
by the National Optical Astronomy Observatory, which is operated by AURA,
Inc., under cooperative agreement with the National Science Foundation.}
task {\tt ellipse} to fit isophotes with a linear step of 2 pixels between 
successive steps, allowing the center, position angle, and ellipticity 
of each ellipse to vary.  The sky background level is determined from the 
best-fit ellipses that have constant surface brightness with respect to 
radius, and the uncertainty on the background is estimated from the 
root mean square fluctuations ($\sigma$) about the mean value.  The limiting 
surface brightness, defined as 3 $\sigma$, is 26.57~$\rm mag~arcsec^{-2}$.
After sky subtraction, the azimuthally averaged one-dimensional (1-D) surface
brightness profile is extracted in logarithmic steps, to increase the S/N in 
the noisier outer regions.  We can clearly see that the profile (right panel 
of Figure~3) at radii larger than $\sim 10\arcsec$ behaves like an 
exponential disk, the scale length of which is 4\farcs21$\pm$0\farcs17 
($2.12\pm0.09$~kpc).  
We do not decompose the inner regions of the galaxy with this data set, as 
the SDSS image lacks sufficient resolution; for that, we turn to the 
high-resolution \hst\ images below.

\subsection{{\it HST}: Optical Image}

The F606W image is our best choice for probing the host galaxy structure 
because (1) the bandpass is relatively red and thus less sensitive to dust and 
young stars, (2) the field of view of WFPC2/PC1 is larger than that of ACS/HRC 
and NICMOS, and (3) its point-spread function (PSF) has less extended wings 
that NICMOS.  A single 500 s exposure was obtained on 21 July 1994 (Proposal 
ID: 5479), with UM~625 placed near the center of the Planetary Camera detector 
(PC1), which has a plate scale of 0\farcs046 pixel$^{-1}$. We remove cosmic 
rays using LA Cosmic\footnote{\tt http://www.astro.yale.edu/dokkum/lacosmic/} 
(van Dokkum 2001), which can detect cosmic ray hits of arbitrary shape and 
size.  Two of the pixels in the center of the galaxy are saturated; we masked 
them out in the analysis below.

Precise sky subtraction is of great importance to obtain accurate photometric 
measurements and structural decomposition.  From the surface brightness 
profile of the SDSS $r$-band image,
\psfig{file=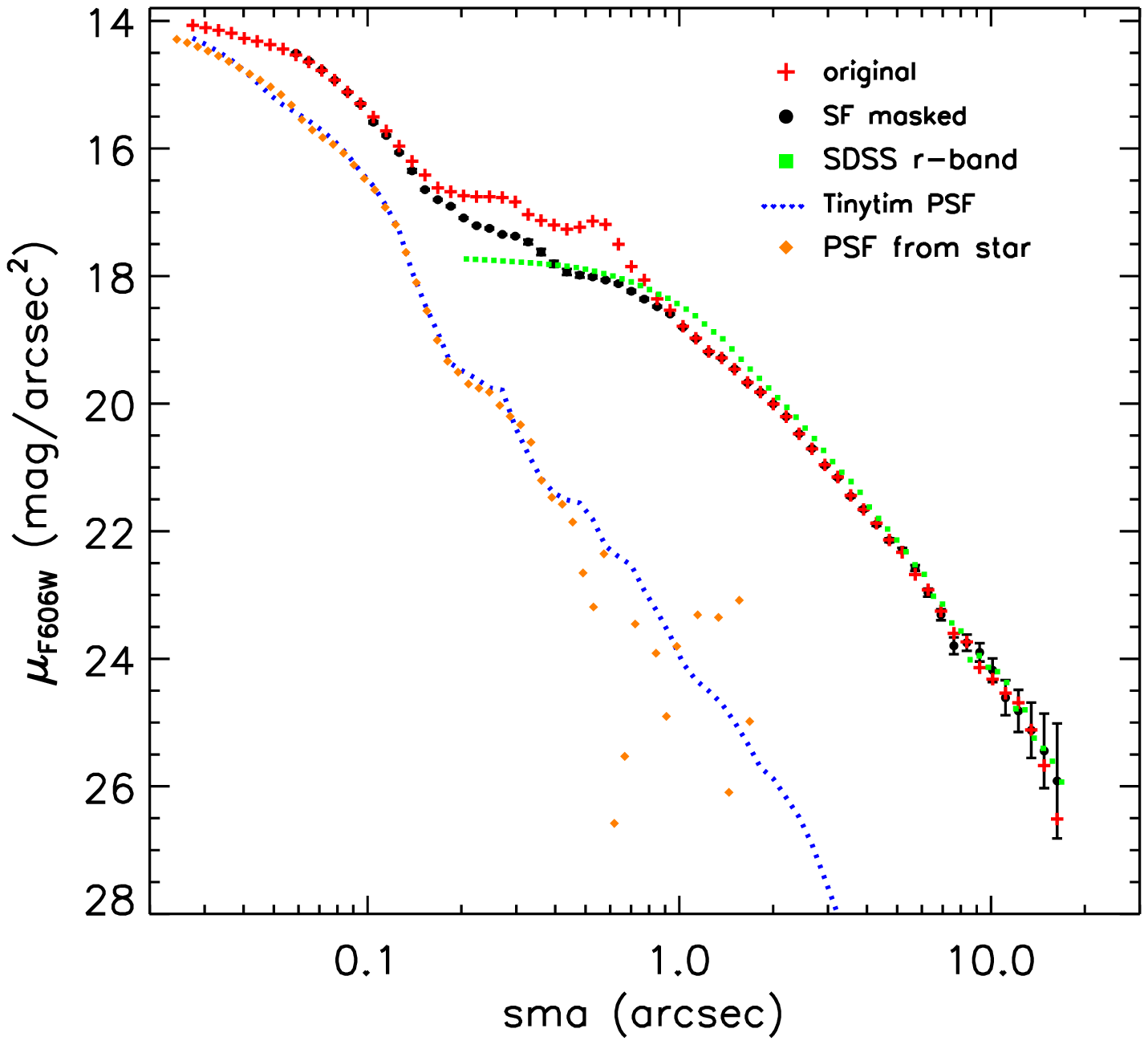,width=9.5cm}
\figcaption[sbp.eps]{
Radial surface brightness profile of UM~625  extracted from the \hst\
WFPC2/PC1 F606W image (red crosses).  The profile with the star-forming ring
masked is displayed with black dots (with $\pm 1\sigma$ error bars).
For comparison we also plot the profile of the PSF created by TinyTim
(blue dotted line) and the PSF profile from a star taken from the WFPC2 PSF
library (orange diamonds).  The PSF profiles are scaled to have the same
central surface brightness as UM~625. The green squares represent the
SDSS $r$-band profile with sky subtracted.
}
\label{f606w-sbp}
\vskip 0.3cm
we know that UM~625 extends close to a 
radius of $20\arcsec$, which means that it fills nearly the entire field of 
PC1 (radius $\sim 18\arcsec$). 
We estimate the background level and its 
uncertainty from the outermost edges of the PC1 chip and from the flanking 
WF2 chips; the limiting surface brightness of the F606W image is 
25.31~$\rm mag\,arcsec^{-2}$.  To examine the effect of the star-forming ring 
on the underlying galaxy structure, we also plot the surface brightness profile 
with the ring region (the region enclosed by the green polygon as indicated in 
Figure~2) masked out.  Figure~4 shows that 
the circumnuclear star-forming region mainly affects the profile on scales of 
0\farcs2$\lesssim r \lesssim$0\farcs8; the outer profile ($r \gtrsim 2\arcsec$)
agrees well with the SDSS $r$-band profile.

Separating the central AGN light from the host galaxy starlight requires 
knowing the PSF to a high accuracy.  Unfortunately there is no bright star in 
the field.  According to the image simulations of Kim et al. (2008), a 
synthetic PSF generated by the TinyTim software (Krist 1995) works reasonably 
well.  We have also searched for empirical stellar PSFs from the WFPC2 PSF 
Library\footnote{http://www.stsci.edu/hst/wfpc2/software/wfpc2-psf-form.html}
and chosen the one closest to the location of UM~625 on PC1.  As shown in 
Figure~4, the stellar PSF agrees well with the TinyTim PSF, 
except in the outer regions where the empirical PSF is much noisier.  For the 
rest of the analysis, we simply adopt the TinyTim PSF.

After background subtraction, we perform a two-dimensional (2-D) decomposition 
of UM~625 using GALFIT (Peng et al.  2002, 2010).  The AGN is represented by 
a point source modeled with the TinyTim PSF, and the galaxy is modeled by 
bulge, fit with a \sersic\ (1968) $r^{1/n}$ function, and a disk, fit with an 
exponential function (equivalent to $n = 1$).   A single-component model for 
the host leaves unacceptably large

\psfig{file=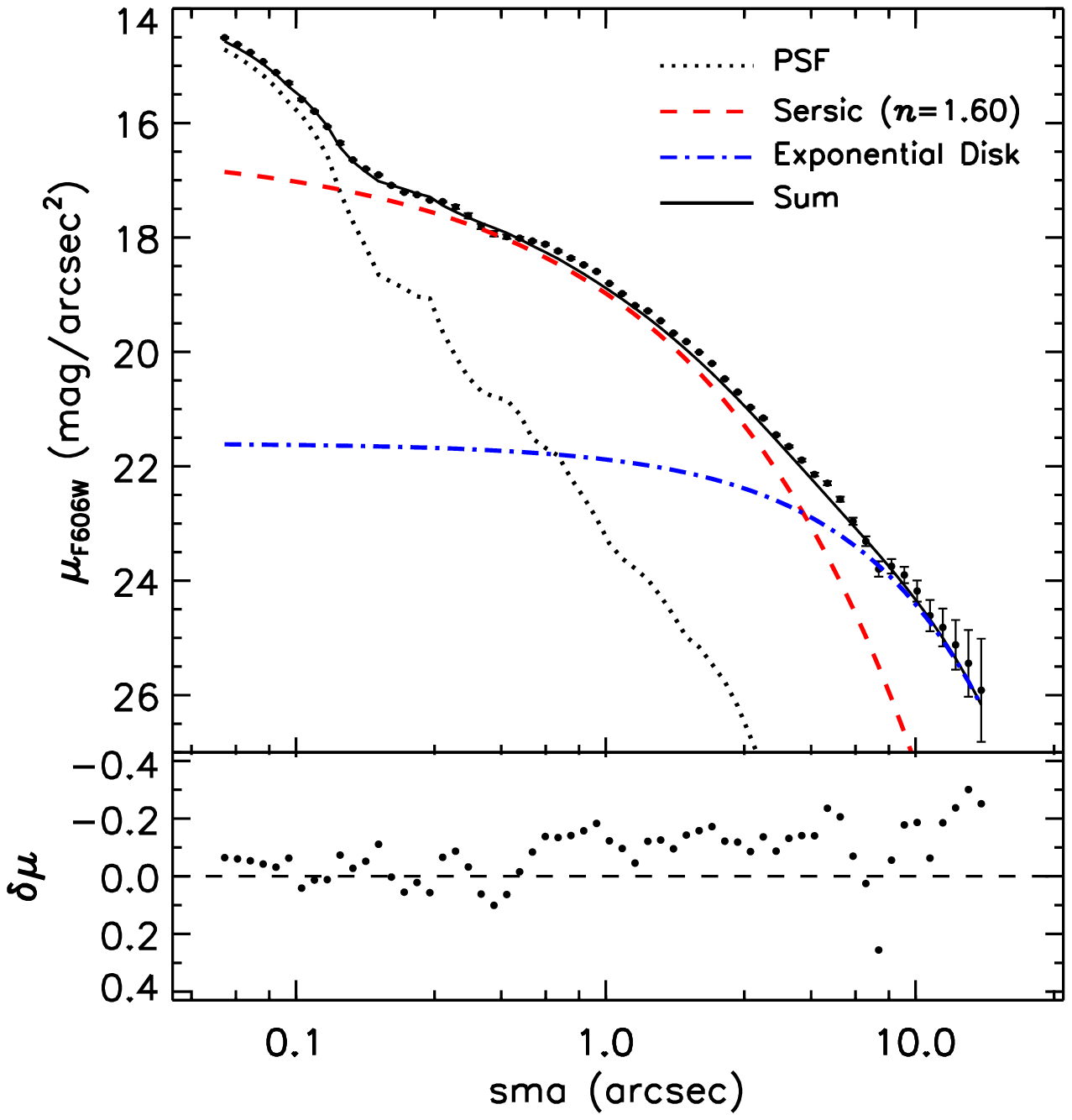,width=9.5cm}
\figcaption[galfit_1d_f606w.eps]{
One-dimensional representation of the three-component 2-D GALFIT model
applied to the WFPC2/F606W image of UM 625: PSF for the nucleus (black dotted
line), $n=1.60$ \sersic\ function for the bulge (red dashed line), and
an exponential function for the disk (blue dot-dashed line).  The sum of
the three components is shown as the black solid line. The observed data are
plotted as black symbols with $\pm 1 \sigma$ error bars.  The bottom panel
shows the residuals between the data and the best-fit model.
}
\label{galfit_1d_f606w}
\vskip 0.3cm
residuals.  The best-fitting two-component 
model (Figure~5) gives a central component with 
\sersic\ index $n = 1.60$ and effective radius $r_e =$ 1\farcs37 (693~pc), 
and a disk with scale length of 3\farcs86 (1.95~kpc), consistent with the 
results from the 1-D decomposition of the SDSS $r$-band image.  The bulge-like
component  is mildly disky, with $c = -0.11$.  
The best-fit parameters are summarized in Table 2.
We identify the central  $n = 1.60$ component with a pseudobulge, because 
pseudobulges generally have $n \lesssim 2$ (Kormendy \& Kennicutt 2004).  
The residual image (top right panel of Figure~2) reveals a faint 
ring-shaped structure on the eastern side of the galaxy, opposite to the 
masked ring region.
\subsection{{\it HST}: NUV Image}
The F330W image was observed on 21 March 2003 using ACS/HRC as part of a study 
on the starburst--AGN connection (Proposal ID: 9379).  ACS is located away 
from the optical axis of \hst, and so it suffers from significant geometric 
distortion that is not corrected by the internal optics. For ease of rejecting 
cosmic rays, the total exposure was divided into two equal exposures of 10 
min each.  These images are then combined using 
{\tt astrodrizzle}\footnote{http://drizzlepac.stsci.edu/}, a new software 
replacing {\tt multidrizzle}, to remove cosmic ray hits and to correct for the 
geometric distortion.  The pixel scale of the combined image is 0\farcs025.
A narrow ring-like star-forming region, ornamented with several bright knots,
encircles nearly half the nucleus.
We begin with creating two synthetic PSFs for each exposure using TinyTim. 
They are centered in the same position as the nucleus
\begin{figure*}
\centerline{\psfig{file=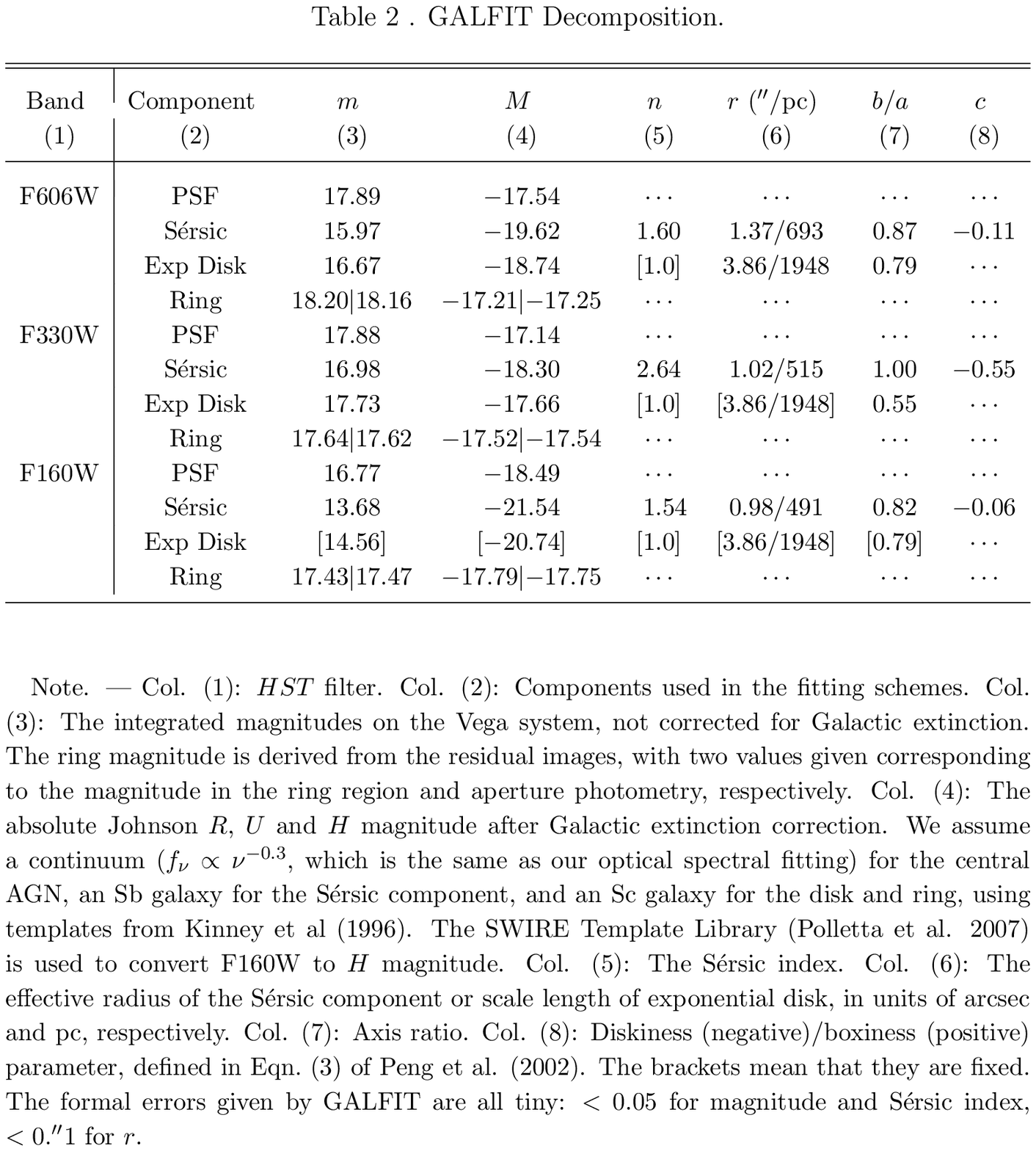,width=12cm,angle=0}}
\end{figure*} 
\vskip 1cm
\noindent
in the HRC image 
to properly reproduce the geometric changes involved in the processing. 
During the modeling, we have also considered the dofocus offset ($\sim6~\mu$m)
given by web-based 
model\footnote{http://www.stsci.edu/hst/observatory/focus/FocusModel.  We did 
not include the offset information for the F606W image because it is only 
available for data taken after 2003.}.  The offset is mostly due to spacecraft 
breathing effects.  The two TinyTim PSFs, still distorted, are almost identical
except for slightly differences due to defocus offsets. We combined them with 
{\tt astrodrizzle} in the same manner as the science images, producing a 
distortion-corrected PSF image.

We decompose the image using GALFIT with the region containing the 
star-forming ring masked out.  Due to the small field of view of the ACS/HRC 
(26\arcsec$\times$29\arcsec), we cannot independently measure the sky 
background; we set it as a free parameter in the fit.  Nor can be obtain any 
meaningful, independent constraint on the parameters of the disk.  Our initial 
attempt to fit the image with a PSF + \sersic\ + disk model failed to converge.
As our primary goal for the F330W image is to measure the brightness of the 
nucleus, we simply set the disk scale length to be identical to that 
obtained from the F606W fit.  The resulting fit yields a bulge with
$n=2.64$ and $r_e =$ 1\farcs02~(515~pc), and a point source flux that is 
very insensitive to the assumptions of the disk component.  The AGN magnitude 
changes by only $\pm 0.01$ mag depending on whether the disk is included or 
not.  If we further force the bulge to have the same $n$ and $r_e$ as 
derived from the F606W fit (i.e. allowing only the luminosity to adjust), the 
resulting AGN magnitude changes by 0.06 mag.

\subsection{{\it HST}: NIR Image}

The NICMOS/NIC1 F160W image was observed on 31 July 1997 (Proposal ID: 7328), 
in three 256~s exposures dithered in an ``L''-shaped pattern.  Each exposure 
was processed with the standard pipeline {\tt calnica} within IRAF/STSDAS.
This task corrects for the nonlinearity of the detector and removes the bias 
value, dark current, amplifier glow, and shading; however, it does not remove 
the ghost ``pedestal'' effect produced by the variable quadrant bias 
(see, e.g., Hunt \& Malkan, 2004).  To correct for this effect, we apply the 
{\tt pedsub} task to the calibrated image, quadrant by quadrant, determine the 
shifts of the dithered images with {\tt xregister}, register them with 
{\tt imshift}, and finally combine them with {\tt imcombine}.

UM~625 extends far beyond the $\sim11\arcsec\times11\arcsec$ field of view of 
NIC1.  In spite of this, a GALFIT model consisting of a PSF + \sersic\ bulge 
can still give a fairly robust measurement of the nuclear magnitude.  During 
the fitting, we set the background free and masked the ring region as before.  
It proved to be impossible to independently measure the disk component at 
all.  Since we know from the optical images that a disk {\it is}\ present, we 
constrain it by fixing its structural parameters to those derived from the 
F606W image and normalizing its luminosity such that $R-H=2$ mag, a value 
typical of disks (e.g., MacArthur et al. 2004). However, we note that none 
of the parameters for the bulge or the AGN point source are significantly 
affected by our assumptions for the disk component.  The best-fit model 

\begin{figure*}[t]
\centerline{\psfig{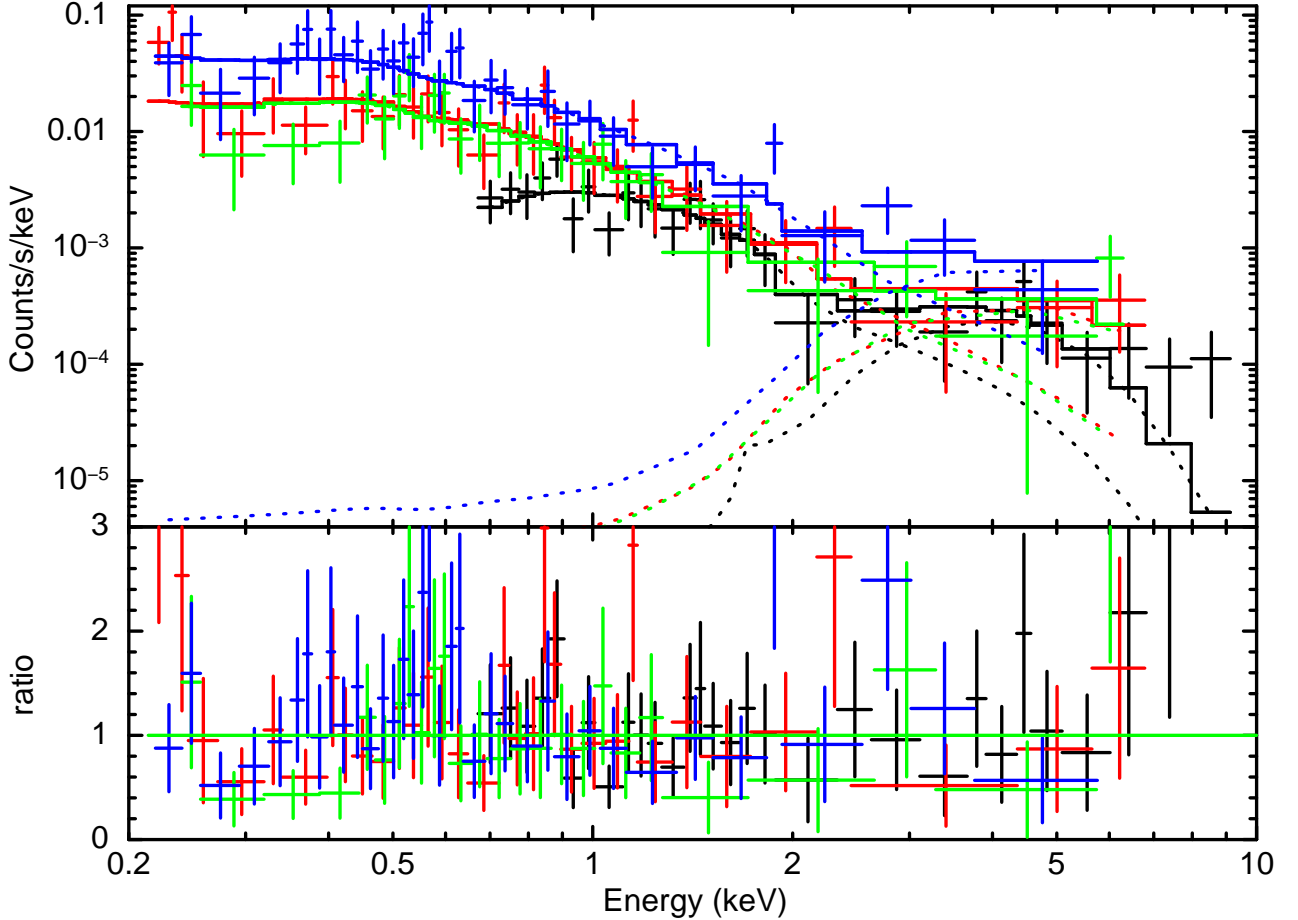}}
\figcaption[xray.eps]{
X-ray spectra of UM~625.  The upper panel shows the data and best-fit model;
the bottom panel shows the ratio between the data and model. 
The dotted lines represent
the two power-law components used in the fitting convolved with the response
of the detectors. Four data sets are shown:
\chandra\ data from 2008 (black) and \xmm\ PN data from 2004 (blue),
2007 (red), and 2008 (green).
}
\label{xray}
\end{figure*}
\noindent
yields $n=1.54$ and $r_e =$ 0\farcs98 (491 pc) for the bulge.  If we fix the bulge 
$n$ and $r_e$ to the values derived from the F606W fit, the AGN magnitude 
changes by 0.21 mag.

\section{Other Multiwavelength Data}

\subsection{X-ray Observations}

UM~625 was observed by \xmm\ on three separate occasions between 2004 and 
2008 (ObsID: 0200430901, 0505930101,  0505930401) and once by \chandra\ in 
2008 (ObsID: 9557).  We used \xmm\ SAS version 12.0.1 and the calibration file 
of June 2012 to reduce the EPIC PN (Str\"uder et al. 2001) archival data.  After 
filtering background flares, the net exposure times are 6.4 ks, 13 ks, and 
13 ks, respectively, for the three PN observations.  The source was detected 
at a large offset angle ($\sim 10\arcsec -11\arcsec$) from the field center 
for the later two PN observations.  To extract the source and background 
spectra,  we defined the source region as a circle with radius $\sim 30\arcsec 
-40\arcsec$  centered at the source position and the background regions as 
three or four circles, far removed from the CCD edges and other sources, with 
radii $\sim 30\arcsec -40\arcsec$ around the source region.  The \chandra\ 
ACIS-S (Garmire et al. 2003) data were reduced with CIAO 4.4.1 (Fruscione et 
al. 2006) and the CALDB 4.4.10 database, adopting standard procedures.  The 
net exposure time was 49~ks.  Because of the large off-axis angle (10\farcs6), 
we extracted the spectra using a 20\arcsec-radius circle for the source and 
four circles around the source region for the background. 
All \xmm\ and \chandra\ spectra were rebinned to have a minimum of 5 
counts in each energy bin after background subtraction, and they were fit 
simultaneously with the same models using {\tt Xspec} (Arnaud 1996).  In view
of the low source counts, we fit the spectra using the Cash-statistic instead 
of $\chi^2$.

We first fit the spectra over the energy range 0.2--10~keV using a single 
absorbed power-law model with the minimum absorption set to the Galactic value 
of $\colnh = 4.04\times10^{20} \, \rm cm^{-2}$ (Kalberla et al. 2005).  The 
fit yields a column density equal to the Galactic value, a photon index 
$\Gamma_{1} = 2.65_{-0.15}^{+0.15}$, and C-statistic = 145 (125 degrees of 
freedom).  The residuals, however, clearly show a flat, hard excess above 3 
keV, which implies that an obscured component exists.  We then added an 
absorbed power-law component with a photon index fixed to $\Gamma_{2} = 1.90$,
typical of AGNs (Shemmer et al. 2008; Constantin et al. 2009).    As shown in 
Figure 6, the fit is improved significantly, resulting in C-statistic = 116 
(123 degrees of freedom).  The soft-band power law, with $\Gamma_1 = 
2.86_{-0.16}^{+0.17}$, is slightly steeper than the previous fit.  The 
absorption column density of the absorbed power-law component is $\colnh= 
9.7_{-5.8}^{+18}\times10^{22} \, \rm cm^{-2}$.  The observed 0.5--2 keV and 
2--10 keV fluxes are $2.5_{-0.2}^{+0.2}\times10^{-14}$ and 
$4.6_{-0.4}^{+0.4}\times10^{-14}$~\flux, corresponding to
$4.0_{-0.3}^{+0.4}\times10^{40}$  and $6.5_{-0.6}^{+0.6}\times10^{40}$~\lum, 
respectively.  Assuming that the obscured power-law component is the intrinsic 
emission from the corona, the unabsorbed 2--10 keV flux is $6.7\times10^{-14}$~\flux, 
corresponding to $9.5\times10^{40}$~\lum.
The monochromatic flux at 2~keV is $7.8\times10^{-32}$~\fnu. 
As the spectra are cut off above 10 keV and no variations are detected 
in each single exposure or among the four exposures, 
we cannot exclude the possibility that the absorption is Compton-thick. 

If we link the spectral slope of the hard component to the soft one, we 
obtain $\Gamma = 2.84\pm0.16$, $\colnh= 14.6_{-7.8}^{+23}\times10^{42} \,\rm cm^{-2}$, 
and a scattering fraction of $10\%$ if we attribute the soft component to 
scattered nuclear emission. We note that this fraction is consistent with 
those commonly reported in other obscured AGNs ($0.1\% - 10\%$; Noguchi et al
. 2010).

\subsection{Ultraviolet Observations}

The Optical Monitor (OM) on {\it XMM-Newton}\ took data for UM 625 in the UVW1 
(2910 \AA) and UVW2 (2120 \AA) filters, each for 1~ks, during the 2004 X-ray 
observations.  Because of the coarse resolution of the OM (PSF $\approx$ 
1\farcs8),
the nucleus is not resolved from the host galaxy, and we consider these global 
(AGN plus host) flux measurements.  We adopt an aperture radius of 10\arcsec\ 
to compute the source flux after estimating the sky background from the region 
between a radius of 15\arcsec\ and 20\arcsec.  The flux is corrected for 
Galactic extinction using the maps of Schlegel et al. (1998) and the reddening 
curve of Fitzpatrick (1999), resulting in $f_{\nu}$(2910 \AA)=$(4.5\pm0.3)
\times10^{-27}$~\fnu\ and $f_{\nu}$(2120 \AA)=$(2.7\pm0.2)\times10^{-27}$~\fnu.

Apart from the {\it XMM-Newton}\ OM data, UM~625 was also imaged 
simultaneously in the near-UV (NUV; 2316 \AA) and far-UV (FUV; 1539 \AA) bands
of \galex\ during its All-sky Imaging Survey (AIS), for a total exposure time
of 112~s on 6 April 2004.  UM~625 was later reobserved during the Medium 
Imaging Survey (MIS) on 18 May 2009, for a total of 1578~s; we adopt the 
MIS data (Bianchi et al. 2012).
After correcting for Galactic extinction, we obtain an NUV flux density of
$f_{\nu}$(2316 \AA)=$(3.17\pm0.04)\times10^{-27}$~\fnu\ and an FUV flux
density of $f_{\nu}$(1539 \AA)=$(1.42\pm0.04)\times10^{-27}$~\fnu. 
The NUV flux density is slightly larger than that of the OM UVW2 band,
whose effective wavelength is only $\sim200$~\AA\ shorter.

\subsection{Radio Observations}

UM~625 was detected by Faint Images of the Radio Sky at Twenty-cm (FIRST; 
Becker et al. 1995) using the Very Large Array in its B configuration.  The 
peak and integrated 20~cm flux density from the FIRST catalog%
\footnote{http://sundog.stsci.edu/cgi-bin/searchfirst} (White et al. 1997)
is 1.89 and 2.69 mJy, corresponding to a monochromatic radio luminosity at 20 cm 
of $2.66\times10^{28}$~and~$3.79\times10^{28}~\rm erg~s^{-1}~Hz^{-1}$. 
These measurements are derived by fitting a 2-D Gaussian function to the 
source, using a map generated from twelve coadded images adjacent to the 
pointing center;  the map has 1\farcs8 pixel$^{-1}$, a  resolution of FWHM = 
5\arcsec, and an rms noise of 0.152 mJy~beam$^{-1}$.
Customarily, the radio-loudness parameter $R$ is defined as the ratio of 
flux densities between 6~cm and 4400~\AA. 
Assuming a radio spectral index $\alpha_r=-0.46$ (Lal \& Ho 2010), 
$f_{\nu}(6~\rm cm)=1.55\times10^{-26}$~\fnu, which, in combination with
$f_{\nu}(4400$~\AA) determined from the SDSS spectral fitting,
gives $R=5.0$.  
Although UM~625 is formally radio-quiet according to the widely used division 
of $R=10$ for radio-loud and radio-quiet AGNs (Kellermann et al. 1989), its 
level of radio activity is still somewhat unusual for low-mass AGNs (e.g., 
Greene et al. 2006).

\subsection{Infrared Observations}

UM~625 is also contained in the \emph{Infrared Astronomical Satellite} 
({\it IRAS}) faint source catalog (Moshir et al. 1990).  The faint source 
catalog contains data for point sources in unconfused regions with flux 
densities typically above 0.2 Jy at 12, 25, and $60\,\mu$m, and above 1.0 Jy 
at $100\,\mu$m.
UM~625 is reliably detected at $60\,\mu$m with a flux density of 
$0.32\pm0.08$~Jy, but only upper limits are given for the other three bands 
(0.13, 0.12, and 0.77~Jy for 12, 25, and $60\,\mu$m, respectively).

\emph{Wide-field Infrared Survey Explorer} ({\it WISE}; Wright et al. 2010), 
the mission most comparable to {\it IRAS} yet with a sensitivity more than 100 
times higher at $12\,\mu$m, has mapped the entire sky in four bands centered 
at 3.4, 4.6, 12, and $22\,\mu$m.  UM~625 is detected with high S/N in all four 
bands, with a flux density, converted from the magnitudes in the {\it WISE}
All-sky Source Catalog, of 4.1, 4.2, 24, and 95~mJy, respectively.

For completeness, we take the following NIR magnitudes from the Two Micron 
All Sky Survey point source catalog (Skrutskie et al. 2006): 
$J=14.25\pm0.05, \, H=13.60\pm0.06$ and $K_{s}=13.24\pm0.05$ mag, which
correspond to a flux density of 3.3, 3.8, and 3.4~mJy, respectively.

\section{Results and Discussion}

\subsection{Black Hole Mass}

With the detection of broad emission lines, we can estimate the mass of the 
central BH using commonly used virial mass estimators for broad-line AGNs.  We 
calculate the virial BH mass using the \ha\ formalism given in Xiao et al. 
(2011; their Equation 6), which is based on Greene \& Ho (2005b, 2007b) but 
updated with the more recent relationship between BLR size and luminosity
of Bentz et al. (2009).  For FWHM(\ha) = 1801 \kms\ and a broad \ha\ luminosity
of $2.4 \times 10^{40}$ \lum, $M_{\rm BH} = 1.6 \times 10^{6}$ \msun, which 
justifies inclusion in the low-mass BH sample of Dong et al. (2012b).

The uncertainty of the above BH mass estimate is not well understood.
As pointed out by Vestergaard \& Peterson (2006), the statistical accuracy of 
the masses from single-epoch virial mass estimators is a factor of $\sim 4$
($1\sigma$), and, for individual mass estimates, the uncertainty can be as 
large as an order of magnitude.  We note that recently Wang et al. (2009) 
recalibrated the BH mass formulas based on single-epoch spectra, stressing the 
nonlinear relation between the virial velocity of the BLR clouds and the FWHM 
of single-epoch broad emission lines.  This nonlinearity probably arises from 
several kinds of nonvirial components incorporated into the total profile of 
broad emission lines in single-epoch spectra (see \S4.2 of Wang et al. [2009] 
for a detailed discussion, as well as Collin et al. [2006] and Sulentic et al. 
[2006]).  The formulas of Wang et al. (2009), calibrated using reverberation
mapping data available to date, which span the mass range from $M_{\rm BH} 
\approx 10^7$ to $10^9$ \msun, does not cover the low-mass regime of interest 
in this paper. Furthermore, we would like to point out that nonvirial components, 
if any, may be less significant in UM~625 than in other AGNs because 
its broad \ha\ profile is roughly symmetrical and close to a Gaussian.

\subsection{Host Galaxy}

The abundant archival \hst\ images covering a wide wavelength baseline provide 
us an excellent opportunity to study the host galaxy of UM~625 with the 
central AGN point source removed.  Our 2-D decomposition of the optical 
(F606W, $R$ band) WFPC2/PC1 shows that the stellar distribution consists of 
two main components: (1) a dominant compact, bulge-like component with 
\sersic\ $n=1.60$, $r_e=693~\rm pc$, and $M_R=-19.62$ mag; (2) an extended 
disk ($n \approx 1$) component with a scale length of $\sim2$~kpc. The disk 
is more evident in the SDSS $r$-band image owing to its deeper limiting 
surface brightness and larger field of view, but its photometric parameters 
are well recovered with the WFPC2/PC1 image alone.  The very limited 
field of view of the ACS/HRC and NICMOS/NIC1 images prevent us from placing 
any meaningful constraints on the disk in the UV and NIR.  We suspect that 
even the structural parameters of the bulge may be partly compromised in these 
bands as a result of the small field size.

Taking the F606W decomposition as reference, the bulge-to-total light ratio 
($B/T$) in the $R$ band is 0.66.  If we use a total host luminosity calculated 
from the SDSS $r$-band image (after subtracting the AGN point source derived 
from the \hst\ decomposition), which may be more reliable because of its 
larger field of view, the bulge-to-total ratio reduces slightly to $B/T = 
0.60$.  Either of these values lies within the range of $B/T$ for S0 galaxies 
(e.g., Simien \& de Vaucouleurs 1986), in agreement with the absence of spiral 
arms in the main disk of the galaxy.  The star-forming ring, which resembles a 
tightly wound spiral, in UM~625 is located in the circumnuclear region 
($\sim 0.5$~kpc) resolved only by \hst\ observations.
Interestingly, the UV-optical and optical-NIR colors 
of the bulge of UM 625 are much bluer than those of typical S0 galaxies; they 
more closely resemble those of an Sb spiral.  This is apparent from comparing 
the observed F330W$-$F606W and F606W$-$F160W colors with synthetic colors 
calculated using {\tt calcphot} in IRAF/SYNPHOT package for various galaxy 
templates (Kinney et al. 1996; Polletta et al. 2007).

In view of its blue color, low \sersic\ index ($n<2$), and disky isophote shape 
($c<0$), the bulge of UM~625 can be convincingly categorized as a pseudobulge, 
commonly present in low-luminosity disk galaxies, including some S0s (see 
Kormendy \& Kennicutt 2004 for a review).  The presence of an ongoing central 
star formation, manifested through the nuclear star-forming ring 
(Section 5.3), further 
supports the pseudobulge interpretation.  This is consistent with previous 
studies, which find that the majority of the host galaxies of low-mass BHs 
with disks are likely to contain pseudobulges rather than classical bulges 
(Greene et al. 2008; Jiang et al. 2011b).  The only possible anomaly is that 
UM 625 has a much higher $B/T$ than usual; for example, the objects in the 
sample of Jiang et al. (2011b) with a detected disk component have an average 
$B/T = 0.23$.

\subsection{Nuclear Star-forming Ring}
As mentioned above, the circumnuclear region of UM~625 contains a blue 
($\rm F330W-\rm F606W=-0.73$ mag, corrected for Galactic extinction) semi-ring 
on scales of $\sim150-400$~pc.  It is most prominent in the NUV (F330W) band, 
which shows a number of bright, compact knots, reminiscent of star-forming 
galaxies with nuclear hotspots (e.g., Barth et al. 1995).  Assuming that the 
UV light mainly arises from young stars, we can use the integrated luminosity 
to estimate the star formation rate. Summing the flux of the residuals above 
the smooth galaxy model within the ring region (green polygon in 
Figure~2) yields a luminosity of 
$L_{\nu}=2.0\times10^{27}$~erg~s$^{-1}$~Hz$^{-1}$ after correcting for 
Galactic extinction, which corresponds to a star formation rate of 0.28 
\msun~yr$^{-1}$ (Kennicutt 1998; his Equation 1).  As a check, aperture 
photometry between radius 0\farcs15 and $1\arcsec$ in the residual image 
yields nearly the same flux, only 0.02 magnitude higher.

Given that the central region of UM 625 experiences significant ongoing star 
formation, as evidenced by the UV light, it is interesting to note that some
of the narrow emission lines, especially the Balmer lines, are inevitably 
contaminated by stellar photoionization.  From Equation (2) of 
Kennicutt (1998), a star formation rate of $\sim 0.3$ \msun~yr$^{-1}$ 
produces an \ha\ luminosity of $3.5 \times 10^{40}$ \lum, which is roughly 
half of the total observed narrow \ha\ emission.  This is nonnegligible.

The star-forming region can potentially contaminate the emission in the 
radio and X-ray bands as well, due to the contribution from high-mass X-ray 
binaries, young supernova remnants, and hot interstellar plasma.  According to 
the empirically calibrated linear relation between star formation rate and 
X-ray luminosity of Ranalli et al. (2003), the ring region produces
$L_{\rm 0.5-2~keV}=1.3\times10^{39}$~\lum; this is only a few percent of the 
observed X-ray luminosity, and so the AGN totally dominates the X-ray emission.
By contrast, the Ranalli et al.'s relation between star formation rate and 
radio emission predicts $L_{\rm 1.4~GHz}=1.1\times10^{28}$~\lum ~ Hz$^{-1}$, 
which is nearly 30\% of the integrated monochromatic radio power detected by 
FIRST.

\subsection{X-ray Spectral Properties}
Although the detection of broad \ha\ and \hb\ qualifies UM 625 as a type 1 AGN,
its X-ray spectrum indicates that it contains a significant intrinsic absorbing 
column density of $\colnh=9.7_{-5.8}^{+18}\times10^{22}\,\rm cm^{-2}$.  
This apparent disagreement between optical and X-ray classification is not uncommon
(e.g., Garcet et al. 2007), as it can arise if the gas responsible for the 
X-ray absorption is highly ionized, instead of neutral, so that the 
accompanying dust would sublimate to yield a much smaller dust-to-gas ratio. 
Indeed, such absorbed type 1 AGNs are usually characterized by either complex 
or warm/ionized absorption (Malizia et al. 2012) arising from ionized gas 
possibly associated with a disk wind (Murray et al. 1995) or ionization cones 
as seen in some objects.

The ratio $T$ between 2--10~keV luminosity to extinction-corrected \oiii\ 
luminosity is also a powerful diagnostic of nuclear X-ray obscuration. 
Previous studies have suggested that objects with $T\leqslant0.1$ are 
invariably Compton-thick, whereas objects with $T\geqslant1$ are almost 
exclusively Compton-thin or unobscured (Guainazzi et al. 2005).  With 
observed $L_{2-10\,\rm keV} = 4.6 \times 10^{40}$ \lum\ and an extinction-corrected
$L_{\rm [O~III]} = 2.3 \times 10^{41}$ \lum, $T=0.2$, placing UM~625 
intermediate between Compton-thick and Compton-thin.

The power-law component for the soft X-ray band is likely different from
the soft X-ray excess commonly seen in type 1 AGNs since it should have been 
largely obscured, if present.  It could arise from a scattered component 
from the nucleus or contamination from the host galaxy; the contribution from 
the star-forming ring is negligible (Section 5.3).  For AGNs with low-mass 
BHs, thermal emission from the accretion disk can also contribute
significantly to the soft X-ray band (e.g., Thornton et al. 2008; Miniutti
et al. 2009).  We have tried to fit our X-ray data using a (disk) blackbody
model, but the data quality is insufficient to reach meaningful conclusions
regarding the nature of the soft X-rays.

\subsection{Spectral Energy Distribution and Bolometric Luminosity}
\begin{figure*}[t]
\centerline{\psfig{file=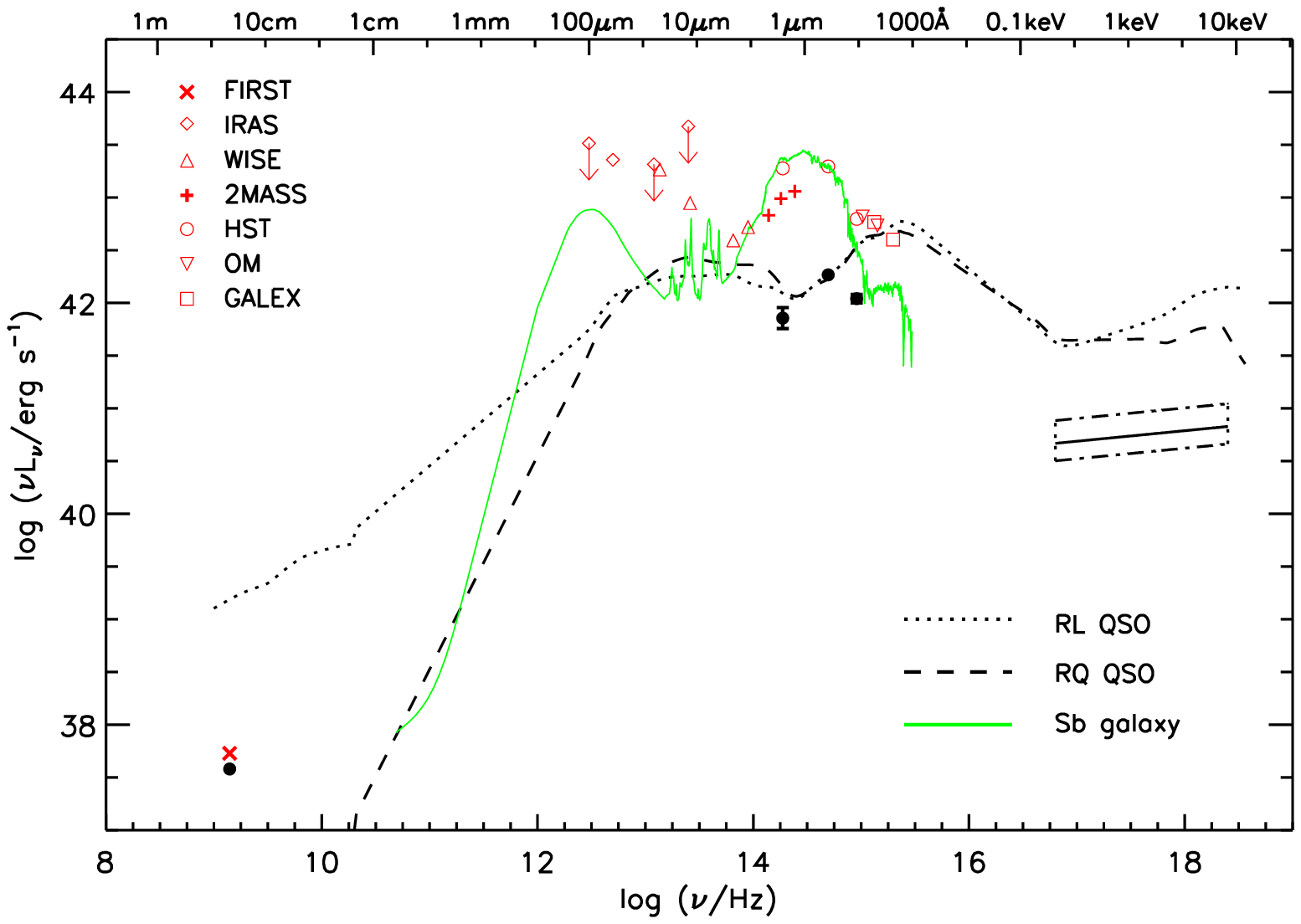,width=18cm}}
\figcaption[sed.eps]{
The SED of UM~625.  The nuclear measurements pertaining to the AGN
are plotted in black dots:  radio data from FIRST (with the contribution from
star-forming ring subtracted), NUV, optical, and NIR data from \hst\
decomposition (error bars denote the range of magnitudes allowed by the
different GALFIT fitting schemes), and unabsorbed X-ray spectrum
from \chandra\ and \xmm\ (1 $\sigma$ uncertainty range given by the dot-dashed
line).  Integrated measurements for the entire host galaxy (plus AGN) are
plotted with various red symbols.
We overplot the median SED of radio-loud (black dotted line) and
radio-quiet (black dashed line) quasars (Elvis et al. 1994),
scaled to the \hst\ nuclear optical point.  We also plot the SED of a
typical Sb galaxy (green line) from the SWIRE template library of
Polletta et al. (2007), scaled  to the \hst\ optical point for the host galaxy.
}
\label{sed}
\end{figure*}
\noindent
We combine all of the photometric data of UM~625 described in Sections 3 and 4 
to construct its broad-band SED (Figure~7); apart from NGC 4395 and 
POX 52, this is one of the most complete SEDs available for low-mass AGNs. The 
median SEDs of radio-quiet and radio-loud quasars (Elvis et al. 1994), scaled 
to UM 625 in the optical band, are overplotted for comparison.  

Even after correcting for absorption, the X-ray emission of UM~625 is still 
somewhat weak compared with the median SED of quasars.  The optical-to-X-ray 
slope \aox\ is $-1.72$, where we adopt the standard definition 
$\alpha_{\rm ox}\equiv 
-0.384{\rm log}[f_{\nu}(2500$~\AA)/$f_{\nu}({\rm 2\,keV})]$ (Tananbaum et al. 
1979) and estimate $f_{\nu}(2500$~\AA) from $f_{\nu}(5100$~\AA) assuming an 
optical-UV continuum spectral index of $-0.44$ (Vanden Berk et al. 2001)%
\footnote{The specific flux at 5100 \AA\ is derived directly from our SDSS 
spectral fit. If, instead, we estimate $f_{\nu}(5100$~\AA) from the \ha\ flux 
(Greene \& Ho 2005b), we obtain $\aox=-1.76$ for the total (narrow plus broad)\ha\ flux and $\aox=-1.53$ for the broad component alone.}.  By contrast,
previous X-ray studies of low-mass AGNs indicate that \aox\ is, on average, 
larger than in high-mass AGNs (e.g., Greene \& Ho 2007a; Miniutti et al. 2009; 
Desroches et al. 2009; Dong et al. 2012a), falling systematically below the 
low-luminosity extension of the \aox-$L_\nu(2500$\AA) relation of Steffen et 
al. (2006).  The previous samples span a wide range in \aox, from 
$\approx -1.7$ to $-1$.  Thus, UM~625 lies among the weakest X-ray sources 
with the lowest \aox.  The origin of the X-ray weakness is not known.
Some may be simply highly absorbed, but others may be intrinsically X-ray 
weak (Dong et al. 2012a).  UM 625 may belong to the latter category.

Low-mass AGNs appear to be exceptionally radio-quiet, if not radio-silent.  In 
a Very Large Array 6~cm survey of the 19 low-mass AGNs from Greene \& Ho 
(2004), Greene et al. (2006) detected radio emission from only one ($\sim 
5$\%), which has $R$=2.8.  This detection rate is approximately the same as 
that found in the larger sample of Greene \& Ho (2007b), whose detected 
sources have $R \approx 1-80$.  UM~625 is clearly detected at 20~cm, at $R = 
5.0$.  Even after correcting for possible contamination from star formation 
(Section 5.3), the source still has $R \approx 3.3$.  While UM 625 is, strictly
speaking, not radio-loud, it is still somewhat unusual in that it belongs to 
the minority of low-mass AGNs that show any radio emission at all.


Despite the fact that the SED of UM 625 still contains many gaps in wavelength 
coverage, it should still offer a more reliable measurement of the bolometric 
luminosity of the system than any estimate based on a single band.
Integrating the median radio-quiet quasar SED of Elvis et al. (1994) after 
scaling it to the nuclear optical point derived from \hst\ for UM 625, 
we obtain \lbol\ = $2.4\times10^{43}$~\lum, which corresponds to \lratio\ = 
0.11 for $M_{\rm BH} = 1.6 \times 10^6$ \msun; had we chosen the radio-loud 
SED template instead of the radio-quiet one, these values would be $\sim$8\% 
higher.  Since the SED of UM 625 does not, in fact, exactly match the 
standard shape of the quasar templates, an alternative approach is to simply 
perform a piecewise power-law integration of the observed points 
(from radio to X-ray).  This yields a bolometric luminosity that is lower 
by a factor $\sim$5, reducing \lratio\ to 0.02.  
Yet a third estimate can be obtained from the strength of the \ha\ 
emission, which scales with the optical continuum emission (e.g., Greene \& 
Ho 2005b), and hence with \lbol\ assuming some canonical bolometric correction 
for the optical band.  Following the formalism of Greene \& Ho (2007b), the 
total (broad plus narrow) \ha\ luminosity of $1.0 \times 10^{41}$ \lum\ 
leads to \lbol\ = $3.2\times10^{43}$~\lum\ and \lratio\ = 0.14. %
\footnote{Dong et al. (2012b) give \lratio\ = 0.04 using the same technique, 
except that they estimate \lbol\ only using the broad component of \ha, which 
is 3 times lower than the narrow component \ha. Here we use total \ha\ while 
the narrow \ha\ is seriously contaminated by the star-forming ring 
as calculated in section 5.3. Our strategy is to give a reliable 
range of \lbol\ and \lratio\ taking all results into consideration.} 
To summarize, UM~625 is estimated to have $\lbol\approx(0.5-3)\times10^{43}$~\lum\
and $\lratio\approx0.02-0.15$, which is between the values for NGC~4395 and POX~52.

\subsection{Black Hole--Bulge Connection}

We end with a short discussion concerning the connection between the mass of 
the central BH in UM 625 and the properties of the bulge of its host galaxy.  
While the SDSS spectrum (Figure 1) does have detected stellar absorption
features, we are not confident that we have resolved them well enough to trust 
the stellar velocity dispersion derived from the continuum fitting (Section 
2).  In lieu of the stars, we use the velocity dispersion of the narrow-line 
gas as traced by the low-ionization lines \sii\ $\lambda\lambda$6716, 6731, 
which are resolved, to estimate the stellar velocity dispersion (Greene \& 
Ho 2005a; Ho 2009).   For $\sigma_{\star} \simeq \sigma_{\rm [S~II]} = 68.1$ 
\kms, the \mbh--$\sigma_{\star}$ relation of G{\"u}ltekin et al. (2009; their 
Equation 7) implies \mbh\ = $1.7 \times 10^{6}$~\msun.  The more recent 
extension of the \mbh--$\sigma_{\star}$ relation to the low-mass end by Xiao 
et al. (2011) yields \mbh\ = $1.3\times10^6\, \msun$.  Both of these estimates 
are in surprisingly good agreement with our virial mass estimate, \mbh\ = 
$1.6\times10^{6}$~\msun, based on the broad \ha\ line (Section 5.1).

By contrast, and in line with other investigations of low-mass AGNs (Greene 
et al. 2008; Jiang et al. 2011a), the BH in UM 625 deviates strongly from the 
BH mass--bulge luminosity relation of inactive galaxies.  Assuming a spectrum 
of an Sb galaxy (Kinney et al. 1996), the F606W magnitude of the bulge 
corresponds to $M_{V}=-19.06$ mag, from which the \mbh--$L_{V,\rm bulge}$ 
relation of G{\"u}ltekin et al. (2009; their Equation 8) predicts $2.2\times 
10^{7}$ \msun, nearly 15 times larger than the virial estimate.

Now, we know that the stellar population of the pseudobulge in UM 625 is 
quite young, most likely significantly younger than the majority of the 
more massive galaxies used to define the \mbh--$L_{\rm bulge}$ relation. 
A proper comparison, therefore, requires that we apply a correction for age or 
mass-to-light ratio to the bulge luminosity of UM 625.  Assuming, as before, 
an Sb galaxy spectrum, for which $B-V\approx 0.7$ mag (Fukugita et al. 1995),
which, in fact, agrees closely with $B-V=0.67$ mag as measured by Salzer 
et al. (1989), a simple stellar population of solar metallicity predicts an age 
of $\sim2$~Gyr and a mass-to-light ratio $M/L_{V}\approx 1.6$ \msun/\lsun\ 
(Bruzual \& Charlot 2003).  For a typical elliptical galaxy, with $B-V\approx 
1.0$ mag, $M/L_{V}\approx 6.3$ \msun/\lsun.  Applying this age correction 
reduces the $V$-band luminosity by about a factor of 4 and the predicted 
BH mass by about a factor of 5.  Although the mass discrepancy is now lower,
the pseudobulge of UM 625, like other pseudobulges containing low-mass 
BHs (Greene et al. 2008; Jiang et al. 2011a), systematically deviate from the 
\mbh--$L_{\rm bulge}$ relation of classical bulges and elliptical galaxies.
Kormendy et al. (2011) suggest that that there are two different modes
of accretion: BHs in classical bulges and ellipticals grow rapidly via 
merger-driven gas infall; in contrast, BHs hosted in pseudobulges grow mainly 
via secular evolution through slower, stochastic processes.

Finally, we note that the pseudobulge of UM~625, as in the hosts of other 
low-mass AGNs (e.g. Greene et al. 2008; Jiang et al. 2011b), deviates from the 
Faber-Jackson relation (Faber \& Jackson 1976) of classical bulges and 
elliptical galaxies, in the sense that it has a lower $\sigma_{\star}$ for a 
given luminosity.  With an $I$-band absolute magnitude of $M_{\rm I}=-20.27$ 
(assuming, as before, a spectrum of an Sb galaxy), the Faber-Jackson relation
of inactive early-type galaxies (Table 5 of Jiang et al. 2011b) predicts 
$\sigma_{\star}=124~\kms$, nearly twice the observed value inferred from 
$\sigma_{\rm [S~II]}$.

\section{Summary}
We conducted a comprehensive, multiwavelength study of the nuclear and 
host galaxy properties of UM~625, a type 1 AGN with a BH mass of 
$1.6\times 10^6$ \msun\ determined through the detection of broad \ha\ 
emission.  

Analysis of \chandra\ and \xmm\ observations reveals a heavily absorbed 
($\colnh= 9.7_{-5.8}^{+18}\times10^{22} \, \rm cm^{-2}$) nucleus with an 
intrinsic X-ray luminosity of $L_{2-10\,\rm keV} = 9.5 \times 10^{40}$ \lum.
The source may be intrinsically X-ray weak ($\aox=-1.72$) compared to higher 
luminosity AGNs.  UM 625 belongs to a minority of low-mass AGNs detected in 
the radio, but it is not strong enough to qualify as radio-loud.  In 
combination with nuclear photometry at UV, optical, and NIR bands extracted 
from high-resolution \hst\ images, the broad-band SED constrains the 
bolometric luminosity to $\lbol\approx(0.5-3)\times10^{43}$~\lum\ and 
$\lratio\approx0.02-0.15$.

We performed a comprehensive analysis of SDSS and \hst\ images to quantify
the structure and morphology of the host galaxy.  The galaxy is an isolated, 
undisturbed, nearly face-on S0 galaxy with a prominent pseudobulge (\sersic\ 
index $n = 1.60$) that accounts for $\sim$60\% of the total light in the $R$ 
band.  The pseudobulge has relatively blue colors ($B-V \approx 0.7$ mag) and
is mildly disky.  Embedded within the central $\sim150-400$~pc is a UV-bright 
semi-ring forming stars at a rate of $\sim$0.3 \msun~yr$^{-1}$.  Consistent 
with other low-mass AGNs, UM 625 follows the \mbh--$\sigma_{\star}$ relation 
but not the \mbh--$L_{\rm bulge}$ relation of inactive galaxies.

\acknowledgments
We thank the referee for a very thorough and helpful review of the paper.
N.J. thanks Minjin~Kim, Zhaoyu Li, Tinggui Wang, Hongguang Shan, Chen Cao, 
and Chien Peng for discussions and for help in image analysis.
We thank Tomas Dahlen and the STScI Help Desk for advice concerning 
the \hst\ photometric system.  We thank Lulu Fan for his early participation 
in this project when he worked with X.-B.D.  
The research of L.C.H. is supported by the Carnegie Institution for 
Science and by NASA grants awarded through STScI.  L.C.H. thanks the Chinese 
Academy of Sciences and the National Astronomical Observatories of China for 
their hospitality while part of this paper was written.  This work is 
supported by the China Scholarship Council, 
Chinese NSF grants NSF-11033007, 
NSF-11133006, 
NSF-11073019, 
a National 973 Project of China (2009CB824800), 
and the Fundamental Research Funds for the Central Universities 
(USTC WK2030220004).
H. Y. and J. X. W. acknowledge support from NSFC (11233002).
Funding for the SDSS and SDSS-II has been provided by the Alfred P. Sloan
Foundation, the Participating Institutions, the National Science Foundation,
the U.S. Department of Energy, the National Aeronautics and Space
Administration, the Japanese Monbukagakusho, the Max Planck Society, and the 
Higher Education Funding Council for England.  The SDSS Web Site is 
http://www.sdss.org/.  This research has made use of the NASA/IPAC 
Extragalactic Database (NED), which is operated by the Jet Propulsion 
Laboratory, California Institute of Technology, under contract with the 
National Aeronautics and Space Administration.

%

\end{CJK*}
\end{document}